\documentclass[a4paper,11pt]{article} 
\pdfoutput=1 
\usepackage{jinstpub} 

\usepackage{caption}
\usepackage{subcaption}
\usepackage{mathrsfs}
\graphicspath{{figures/}{/plots}} 

\usepackage{xspace}

\newcommand {\ie}{\mbox{i.e.}\xspace}     
\newcommand {\eg}{\mbox{e.g.}\xspace}     

\title{Gain and time resolution of 45\,$\mu$m thin Low Gain Avalanche Detectors before and after irradiation up to a fluence of $10^{15}$\,n$_{eq}$/cm$^2$}

\author[a,1]{J.~Lange\note{Corresponding author.},} 
\author[b]{M.~Carulla,}
\author[a]{E.~Cavallaro,}
\author[c]{L.~Chytka,}
\author[d]{P.M.~Davis,}
\author[b]{D.~Flores,}
\author[a]{F.~F\"{o}rster,}
\author[a,e]{S.~Grinstein,}
\author[b]{S.~Hidalgo,}
\author[c]{T.~Komarek,}
\author[f]{G.~Kramberger,}
\author[f]{I.~Mandi\'{c},}
\author[b]{A.~Merlos,}
\author[c]{L.~Nozka,}
\author[b]{G.~Pellegrini,}
\author[b]{D.~Quirion,}
\author[g]{T.~Sykora}

\affiliation[a]{Institut de F\'{i}sica d'Altes Energies (IFAE), The Barcelona Institute of Science and Technology (BIST), 08193 Bellaterra (Barcelona), Spain}

\affiliation[b]{Centro Nacional de Microelectronica (CNM-IMB-CSIC), Campus UAB, 08193 Bellaterra (Barcelona), Spain}
\affiliation[c]{Palack\'{y} University, SLO\&RCPTM, Olomouc, Czech Republic}

\affiliation[d]{Centre for Particle Physics, Department of Physics, University of Alberta, Edmonton, AB T6G 2G7, Canada}

\affiliation[e]{ICREA, Pg. Llu\'{i}s Companys 23, 08010 Barcelona, Spain}
\affiliation[f]{Jozef Stefan Insititute, Jamova 39, SI-1000 Ljubljana, Slovenia}

\affiliation[g]{Charles University in Prague, Faculty of Mathematics and Physics, Institute of Particle and Nuclear Physics, V Holesovickach 2, CZ - 18000 Praha 8, Czech Republic}

\emailAdd{joern.lange@cern.ch} 

\abstract{ 

Low Gain Avalanche Detectors (LGADs) are silicon sensors with a built-in charge multiplication layer providing a gain of typically 10 to 50. Due to the combination of high signal-to-noise ratio and short rise time, thin LGADs provide good time resolutions. 

LGADs with an active thickness of about 45\,$\mu$m were produced at CNM Barcelona. Their gains and time resolutions were studied in beam tests for two different multiplication layer implantation doses, as well as before and after irradiation with neutrons up to $10^{15}$\,n$_{eq}$/cm$^2$.

The gain showed the expected decrease at a fixed voltage for a lower initial implantation dose, as well as for a higher fluence due to effective acceptor removal in the multiplication layer. Time resolutions below 30\,ps were obtained at the highest applied voltages for both implantation doses before irradiation. Also after an intermediate fluence of $3\times10^{14}$\,n$_{eq}$/cm$^2$, similar values were measured since a higher applicable reverse bias voltage could recover most of the pre-irradiation gain. At $10^{15}$\,n$_{eq}$/cm$^2$, the time resolution at the maximum applicable voltage of 620\,V during the beam test was measured to be 57\,ps since the voltage stability was not good enough to compensate for the gain layer loss. The time resolutions were found to follow approximately a universal function of gain for all implantation doses and fluences.

}


\arxivnumber{1703.09004} 

\begin{document}

\maketitle 
\flushbottom 

\section{Introduction}
\label{sec:intro}

Segmented silicon detectors play a dominant role today as precision tracking devices in high-energy-physics (HEP) experiments with typical spatial resolutions in the order of 10\,$\mu$m. Recently, there has been large interest in studying and developing also the timing capabilities of silicon detectors. The intermediate studies concentrate on optimising spatial and timing performance separately, but concepts of 4D tracking (high precision space and time measurements combined in one device) are also being conceived~\cite{bib:UFSDproposal,bib:4Dtracking}. An important application of timing with an overall precision of 10--30\,ps could be \eg pile-up removal at the Large Hadron Collider (LHC) experiments since the exact knowledge of the time-of-arrival of a particle hints at its originating primary vertex. This is needed for forward detectors such as the ATLAS Forward Proton (AFP) experiment~\cite{bib:AFPreference1} or the CMS-TOTEM Precision Proton Spectrometer (CT-PPS)~\cite{bib:CTPPS,bib:CTPPStiming}, which require 10--20\,ps timing detectors already in 2017, or for high-luminosity (HL-)LHC upgrade projects such as the ATLAS High Granularity Timing Detector (HGTD)~\cite{bib:HGTD}, which are planned to be installed around 2024.

The time resolution of silicon detectors has two major contributions (neglecting contributions from the digitisation of the signal or non-uniform weighting fields): the jitter from the noise of the sensor-electronics system, which can be improved with a high signal-to-noise ratio (S/N) and a short rise time of the signal; and charge deposition variations, sometimes called Landau noise. The latter leads on the one hand to amplitude variations, which can be typically taken into account using amplitude corrections or constant-fraction thresholds. On the other hand, however, local charge deposition variations in different depths of the sensor lead to uncorrectable intrinsic signal fluctuations. Their impact can be mitigated by using thinner sensors~\cite{bib:Lorenzo,bib:LGADDesignOpt}.

Standard silicon detectors without internal gain and thicknesses between 100 and 200\,$\mu$m have been shown to provide a time resolution of 100--150\,ps~\cite{bib:Lorenzo, bib:NA62TB} for a minimum-ionising particle (MIP). Since a high signal-to-noise ratio is key to obtain a good time resolution, silicon detectors with a built-in charge multiplication mechanism providing internal gain can provide advantages. Early tests on commercial silicon avalanche diodes with active thickness of 140\,$\mu$m and a gain of 50 achieved time resolutions down to 65\,ps~\cite{bib:APD}. Recently, following the observation of charge multiplication as a radiation effect in highly irradiated silicon detectors~\cite{bib:CMJoern,bib:CMIgor,bib:CMGian}, a new technology called Low Gain Avalanche Detectors (LGAD)~\cite{bib:LGAD} has been developed at CNM Barcelona in the framework of the CERN-RD50 collaboration~\cite{bib:RD50}. It is based on implanting a highly doped p-type layer between the high resistivity p-type bulk and the $n^+$ implant, which acts as high-field charge multiplication layer with a target gain of 10--50 and custom segmentation between the $\mu$m and mm levels. The first devices with a thickness of 300\,$\mu$m and a gain of about 10 showed already a time resolution of 120\,ps for a MIP~\cite{bib:UFSD300umTB}. It was predicted that thinner detectors would improve the timing performance significantly to about 30\,ps for 50\,$\mu$m thin devices with a gain in the order of 10, due to both a steeper signal slope and less Landau noise compared to thicker LGADs~\cite{bib:UFSD300umTB}.

In May 2016, CNM finished the first production of LGADs with an active thickness of about 45\,$\mu$m and a gain of larger than 10. The first timing studies of these devices have been performed in the AFP beam test in June/July 2016 at the CERN-Super Proton Synchrotron (SPS) with 120\,GeV pions, which will be reported here. Also other groups studied the timing performance before irradiation~\cite{bib:UFSD50umTBNicolo,bib:HGTD}.

However, a key to the above-mentioned applications in LHC experiments is radiation hardness. For example, AFP is exposed to a highly non-uniform irradiation with a peak fluence of about $10^{15}\,n_{eq}$/cm$^2$ per year~\cite{bib:AFPreference1}, similarly for CT-PPS. Also HGTD is required to withstand radiation levels in the same order of magnitude. However, in previous studies on 300\,$\mu$m thick LGADs, it was found that the gain is highly reduced after irradiation due to the effective removal of the initial acceptors~\cite{bib:LGADradiationGregor}. Hence, a focus of this study is the gain and timing performance of 45\,$\mu$m LGADs after irradiation with neutrons to fluences of $3\times10^{14}$ and $10^{15}\,n_{eq}$/cm$^2$, which was measured in the AFP beam test in September 2016 at the CERN-SPS. The first results of LGAD time resolutions after irradiation will be reported in this paper.



\section{Technology and samples}
\label{sec:samples}

\begin{figure}[hbtp]
	\centering
	\includegraphics[width=14cm]{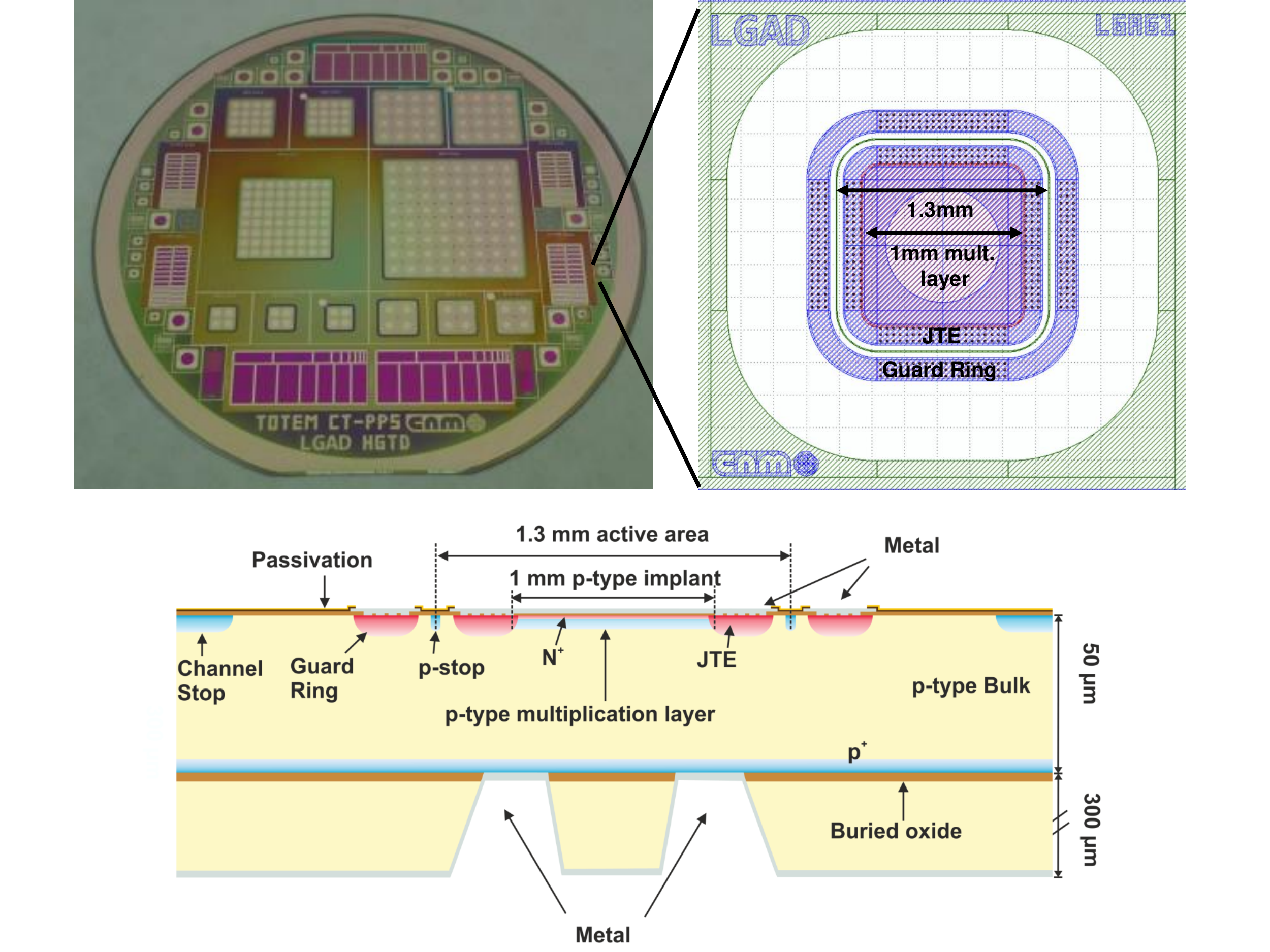}
	\caption{Picture of the wafer (top left), as well as top view (top right) and side view (bottom) of the LGA pad diode structure.}
	\label{fig:samples}
\end{figure}

LGAD devices were produced by CNM Barcelona on 4" silicon-on-insulator (SOI) wafers with nominally 50\,$\mu$m thickness on a 300\,$\mu$m thick support wafer and 1\,$\mu$m buried oxide\footnote{The first run 9088, which is studied here, finished in May 2016.}. Due to the diffusion of the highly doped n$^+$ and p$^+$~implants at the front and back side, respectively, the active thickness is reduced to about 45\,$\mu$m, which is consistent with capacitance measurements (see section~\ref{sec:lab}). Figure~\ref{fig:samples} (top left) shows a picture of the wafer with different structures. It mainly includes single pad diodes of overall active area of 1.3$\times$1.3\,mm$^2$ (LGA) and 3.3$\times$3.3\,mm$^2$ (LGB) in the periphery, as well as segmented arrays of diodes with various dimensions in the centre of the wafer, which were designed for the HGTD and CT-PPS/TOTEM experiments. 

In this study, the small single pad diodes LGA were used. Figure~\ref{fig:samples} shows a top and side view of it. The bulk is p-type doped with 12\,k$\Omega$cm resistivity and the central pad is made of a 1.2\,mm wide n$^+$~implantation to provide the p-n junction, including a junction termination extension (JTE) at the edges to improve the break down behaviour. It is surrounded by a p-stop implant, defining the 1.3\,mm wide active area, and a guard ring. The gain is provided by a 1.0\,mm wide central p-type multiplication layer underneath the n$^+$~implant, which enhances the electrical field in this region to values sufficient for impact ionsiation. Three different multiplication layer implantation doses\footnote{In the following simply referred to as \emph{dose}, not to confuse with the irradiation level which will be referred to in terms of particle \emph{fluence}.} were used on different wafers in order to provide devices with different gain and break down voltage ($V_{BD}$) behaviour: 1.8 (low), 1.9 (medium) and 2.0 (high) $\times10^{13}$\,cm$^{-2}$. In this paper, the low and medium implantation doses were studied. The top implant is metalised with aluminium with a central hole for light injection. Four holes are etched through the insensitive substrate from the back side, which is subsequently covered with aluminium, to provide the back side contact to the thin active area.

Two of the medium-dose samples were irradiated with neutrons at the TRIGA reactor in Ljubljana to a 1-MeV-neutron-equivalent fluence of $3\times10^{14}$\,n$_{eq}$/cm$^2$, and another two to $10^{15}$\,n$_{eq}$/cm$^2$.

An overview on the samples studied is given in table~\ref{tab:samples}. Typically, for each type of dose and fluence, there are two copies of LGAD diodes called briefly L1 and L2 (for the low dose, also a third device L3 was used, but only for the Sr90 measurement). In the following, the devices will be referred to according to their short names listed in the table.

\begin{table}[hbt]
			\centering
			\caption{Overview on samples and measurements.}
			\label{tab:samples}
			\begin{tabular}{|c|c|c|c|c|c|c|c|c|}
			
			\hline		
 Sample & Dose [$10^{13}$\,/cm$^2$] & Fluence [$10^{14}$\,n$_{eq}$/cm$^2$]& Measurements & Short Name \\
\hline
W3-LGA-61 & 1.8 & 0 & IV,CV,Beam & low,unirr,L1 \\
W3-LGA-71 & 1.8 & 0 & IV,Beam & low,unirr,L2 \\
W3-LGA-33 & 1.8 & 0 & Sr90        & low,unirr,L3 \\
\hline
W5-LGA-45 & 1.9 & 0 & IV,CV,Sr90,Beam & med,unirr,L1 \\
W5-LGA-81 & 1.9 & 0 & IV,Sr90,Beam & med,unirr,L2 \\
\hline
W5-LGA-51 & 1.9 & 3 & IV,Sr90,Beam & med,3e14,L1 \\
W7-LGA-45 & 1.9 & 3 & IV,Sr90,Beam & med,3e14,L2 \\
\hline
W5-LGA-43 & 1.9 & 10 & IV,Sr90,Beam & med,1e15,L1 \\
W7-LGA-35 & 1.9 & 10 & IV,Sr90,Beam & med,1e15,L2 \\
			\hline 
			\end{tabular} 
	\end{table}
	
\section{Laboratory characterisation: IV and gain}
\label{sec:lab}

\begin{figure}[hbtp]
	\centering
	\includegraphics[width=7.5cm]{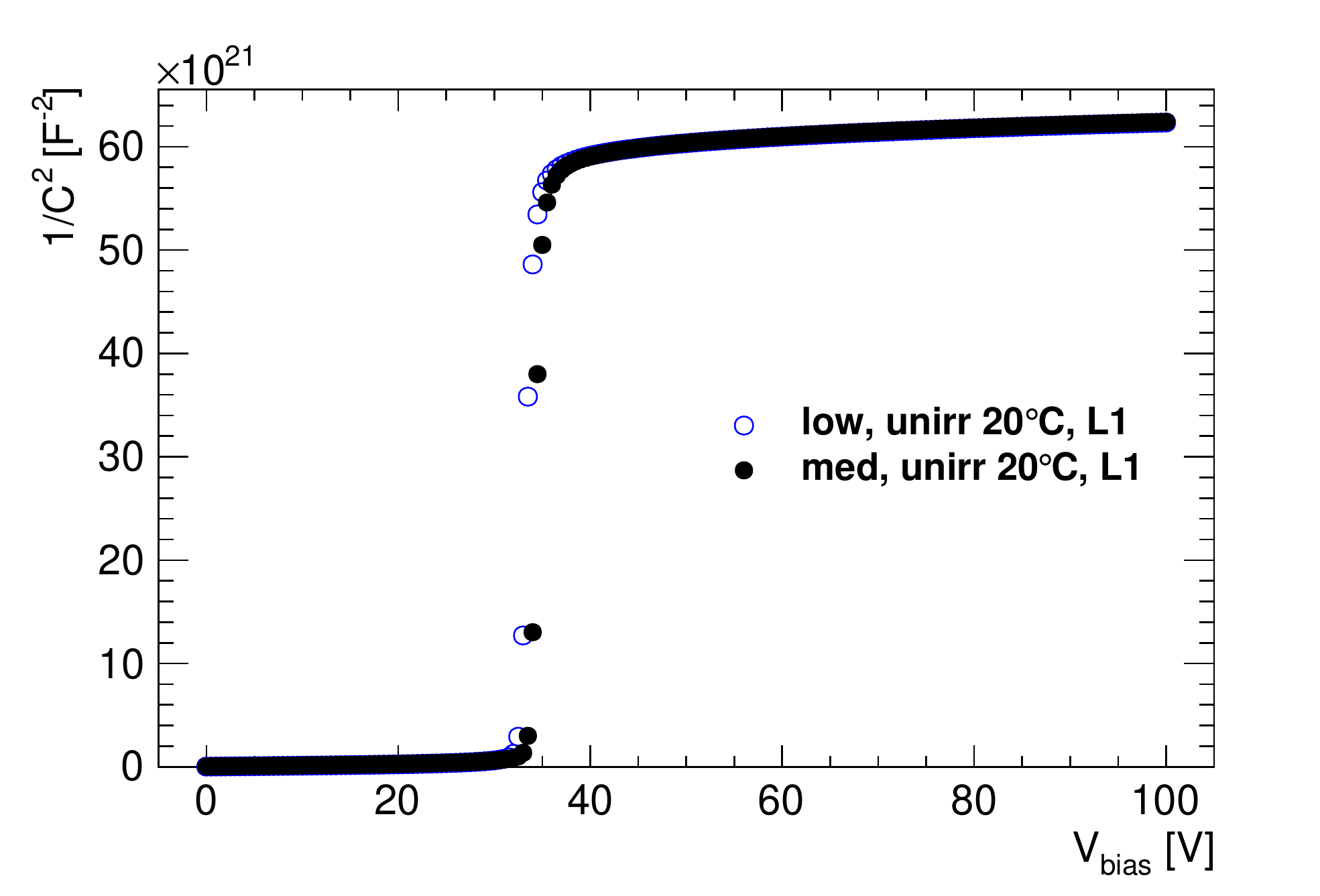}
	\includegraphics[width=7.5cm]{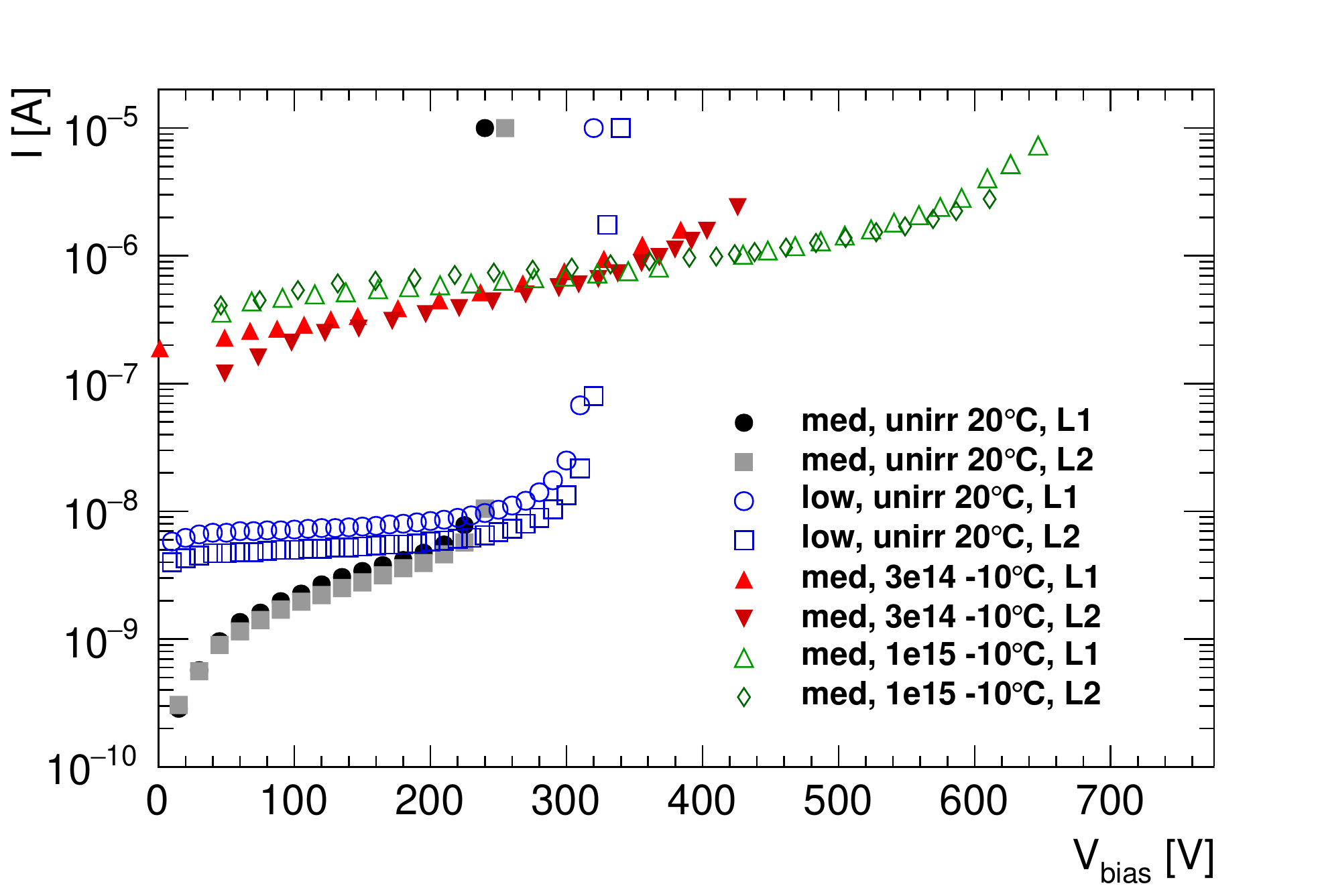}
	 \includegraphics[width=7.5cm]{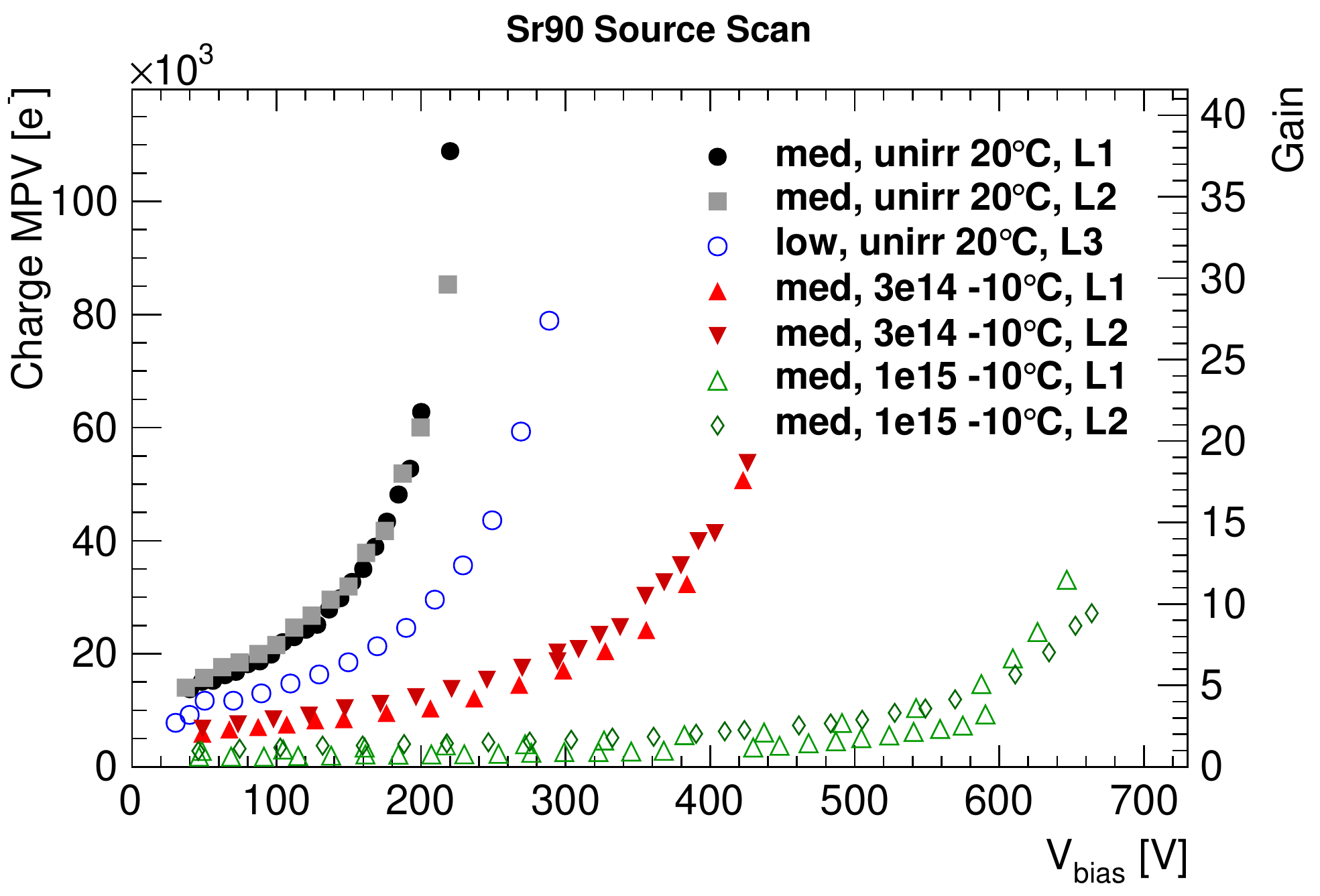}
	\caption{The $1/C^2$ vs. voltage curve (top left) and current-voltage (IV) characteristics (top right). Bottom: Charge collected and corresponding gain as a function of bias voltage measured with Sr90 beta particles.}
	\label{fig:lab}
\end{figure}

Before the beam tests, the devices were characterised in the laboratories of CNM, IFAE Barcelona and JSI Ljubljana with current-voltage (IV), capacitance-voltage (CV) and Sr90 charge collection measurements.

Figure~\ref{fig:lab} (top left) shows the $1/C^2$ vs. voltage curves for one device with low and medium implantation dose before irradiation, respectively. Up to 32\,V (for the medium dose about 1\,V later due to the higher doping), the thin highly-doped p-type multiplication layer is slowly depleting, and $1/C^2$ stays at a low level. Then within about 3\,V the remaining high-resistivity bulk is fully depleting. The capacitance of the LGAD devices was measured to be 3.9\,pF after full depletion, consistent with expectations for a 45\,$\mu$m active thickness.

Figure~\ref{fig:lab} (top right) shows the IV curves for all devices, measured at room temperature before and at -10$^{\circ}$C after irradiation. Before irradiation, the current is approximately at the nA level after full depletion at about 40\,V. The current never reaches a real plateau and keeps increasing due to the charge multiplication (note the logarithmic scale), before it reaches a hard break down at about 240\,V (320\,V) for the medium (low) dose. After irradiation, the current increases with fluence, as expected, and reaches the $\mu$A level (for -10$^{\circ}$C). The current increase with voltage is stronger for the lower fluence, and also the break down is reached earlier since, as will be shown below, the gain degrades with fluence.

The charge in response to beta particles from a Sr90 source was measured with the setup in Ljubljana described in detail in reference~\cite{bib:EPIcharge}. The devices were mounted in aluminium boxes that were placed between the source and a scintillator trigger. The signal was processed with a charge-sensitive amplifier and a custom-made 25\,ns shaper, recorded with an oscilloscope and analysed offline. The charge was calibrated with an Am source. In the following the most-probable value (MPV) of a Landau-Gauss fit to the charge spectrum will be used. The gain is calculated as the ratio of the charge of the LGAD devices to the charge of an equivalent 45\,$\mu$m thin pad diode from the same run, but without p-type charge multiplication implant. This no-gain reference charge was measured to be 2.88\,ke$^{-}$, consistent with expectation~\cite{bib:PDG}. Figure~\ref{fig:lab} (bottom) shows the measured charge and corresponding gain as a function of voltage. The two devices of the same implantation dose or fluence behave very similarly. At a fixed voltage, the gain is highest for the medium dose before irradiation, which reaches \eg a gain of 20 at 200\,V, whereas the low dose reaches a lower value as expected, namely only 10. However, since the voltage stability of the low-dose device surpasses the one of the medium-dose one, similar end-point gains can be also reached with lower doses, but at higher applied voltages. It should be noted that the exact end point is not well defined. The measurements were stopped when the waveform started to show instabilities like micro discharges or baseline fluctuations. This was typically the case close to the hard current break down and highly depends on the setup and operating and environmental conditions. For example in the beam tests, the measurements could be typically performed up to slightly higher voltages (see section~\ref{sec:beamTest}). Due to the large slope at high voltages, this can make a considerable difference in the gain achieved. After irradiation, the gain is highly reduced at the same voltage, as it has been already observed in thicker LGAD devices~\cite{bib:LGADradiationGregor}. This can be explained by an effective removal of the initial acceptors of the p-type multiplication layer. For example at 200\,V, the gain is only 4 at $3\times10^{14}$\,n$_{eq}$/cm$^2$ and about 1 at $10^{15}$\,n$_{eq}$/cm$^2$. But also here, the improved voltage stability after irradiation helps to recover at least part of the original gain. A maximum gain of about 18 (10) is reached at 425\,V (650\,V) at $3\times10^{14}$\,n$_{eq}$/cm$^2$ ($10^{15}$\,n$_{eq}$/cm$^2$). A more detailed and extensive study of radiation effects in 45\,$\mu$m thin LGADs is under way and will be published separately. It is observed that the effective acceptor removal of the multiplication layer seems to be completed at a fluence of about $2\times10^{15}$\,n$_{eq}$/cm$^2$, where the LGAD sensors behave in the same way as standard reference diodes without built-in multiplication layer~\cite{bib:GregorTrento}. However, both LGADs and standard reference sensors do show charge multiplication at such high fluences with a gain up to 8, which in that case originates from high fields from radiation-induced charged bulk defects as observed before in other thin silicon pad diodes~\cite{bib:CMJoern}.

\section{Beam test measurements}
\label{sec:beamTest}

The first timing measurements of 45\,$\mu$m LGAD detectors before and after irradiation were performed in the AFP beam tests in June/July and September 2016 in the H6B beam line of the CERN-SPS North Area with 120\,GeV pions.

\subsection{Beam test setup and operation}
\label{sec:setup}

\begin{figure}[hbtp]
	\centering
	\includegraphics[width=12cm]{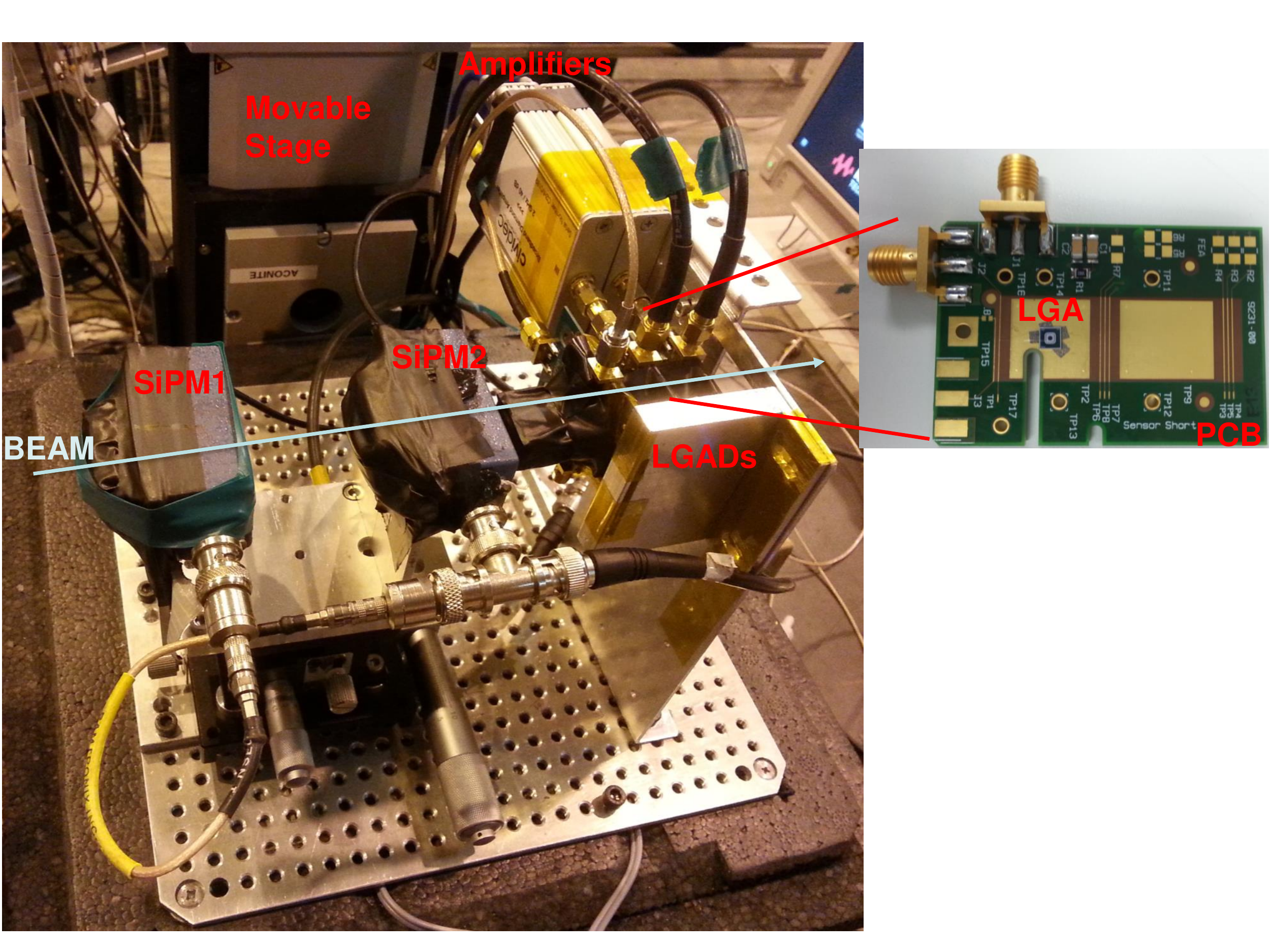}
	\caption{The LGAD measurement setup during the AFP beam test in the SPS H6B beam line.}
	\label{fig:setup}
\end{figure}

Figure~\ref{fig:setup} shows a picture of the beam test setup. The LGAD detectors were mounted on a Printed Circuit Board (PCB) developed originally for Transient Current Technique measurements. The bias voltage was applied via a low-pass RC filter to the back side. The front side was connected via a bond wire and SMA connectors to an amplifier, typically a broad band CIVIDEC C2 TCT model with a band width of 2\,GHz~\cite{bib:CIVIDEC}. The guard ring was not connected. The signal was recorded by an oscilloscope, typically the Agilent infiniium DSA91204A oscilloscope with 40\,GS/s sampling rate and a band width of 12\,GHz, which was however reduced online to 1\,GHz by default (see discussion below). 
Optionally, it was also possible to use an AFP Constant Fraction Discriminator (CFD) as described in~\cite{bib:AFPbeamTests} between the amplifier and the oscilloscope. As fast 10\,ps timing references, two devices based on Cherenkov-light emitting quartz bars coupled to silicon photomultipliers (in the following simply referred to as SiPM1 and 2) were used~\cite{bib:AFPbeamTests}. The quartz bars have a cross section of $3\times3$\,mm$^{2}$ and are 30\,mm long in beam direction. A bias voltage of 30.7\,V was applied to the SiPMs and the signal was discriminated with the CFDs before being recorded by the oscilloscope. The SiPMs were placed on a mechanically adjustable table to align them with respect to the LGADs. The whole setup was mounted on a base plate connected to a remotely controllable micrometer precision table. The base plate could be covered with a styrofoam box for light-tightness and thermal insulation during cooling. The unirradiated devices were measured without cooling at room temperature. The cooling of that setup during the measurements of the irradiated devices was performed with dry ice. The on-sensor temperature was extracted using the temperature dependence of the leakage current by comparing the current during the measurement to the one measured at different temperatures in a laboratory climate chamber. In this way, the on-sensor temperature during the measurements with the $3\times10^{14}$\,n$_{eq}$/cm$^2$ ($10^{15}$\,n$_{eq}$/cm$^2$) devices was determined as -6$^{\circ}$C (-15$^{\circ}$C). For the $3\times10^{14}$\,n$_{eq}$/cm$^2$ devices, an additional run was performed in a beam-test compatible climate chamber set to -20$^{\circ}$C. In that case, the on-sensor temperature was found to be consistent with the set one.

At the beginning of the beam test, setup variations and operation conditions were studied to find the optimal measurement configuration, as well as to understand the running stability. Besides the CIVIDEC C2 TCT amplifier, also other broad band amplifiers were tested such as the AFP custom-made pre-amplifiers~\cite{bib:AFPbeamTests} or the Particulars TCT amplifier~\cite{bib:Particulars}, but both were found to give 5--10\,ps worse time resolution due to a lower S/N. The performance of the CIVIDEC C6 charge-sensitive amplifier was even much worse (about 100\,ps) due to the non-optimised shaping time of 4\,ns, much longer than the original 500\,ps rise time of the signal (see section~\ref{sec:waveformParameters}). Also another oscilloscope (LeCroy, 2\,GHz, 20\,GS/s) was studied and, moreover, the sampling rate of the Agilent oscilloscope was reduced online to 10 or 20\,GS/s, but these variations were found to have negligible impact. However, the setting of the vertical scale of the oscilloscope was found to contribute to the noise and hence the time resolution due to the rather large
digitisation uncertainty introduced by oscilloscopes with 8-bit vertical resolution. It was set by default to 50\,mV/div and kept unchanged during the measurements to obtain comparable results. Also the band width of the oscilloscope, which could be set online, was found to impact the time resolution, since it should be large enough not to deteriorate the fast rise time of the signal, but small enough to filter out high-frequency noise. This was studied in the range between 0.5 and 12\,GHz and the optimum was found at 1\,GHz, which was set as default. This corresponds to a rise time of 340\,ps for a step function, which is below the intrinsic pulse rise time of about 500\,ps.

Due to ambient noise in the beam area and the non-optimal assembly with long wires and connectors between the sensor and the amplifier, run-to-run variations were observed. The noise was found to vary between 3 and 4\,mV, and the amplitude and integrated charge by up to 30--50\%. However, the impact on the final time resolution was found to be typically only about 10\%.

Some runs were taken with two SiPMs and two LGAD devices connected to the oscilloscope channels, but in some cases only one SiPM was available, or, in the case of the additional run in the climate chamber, no SiPM. Typically 10,000 events were recorded for each run. The trigger could be basically selected to be any channel, and for test run periods, data with different triggers (L1, L2 and SiPM2) were taken and compared. Since no significant difference or bias depending on the trigger channel was found and the purity of the data sample was higher when triggering on the LGADs themselves, this was taken as the default trigger option. The LGAD trigger level for low- and medium-voltage runs was typically set to 15--20\,mV, which was high enough above the noise but low enough not to cut into the signal. At the highest measured voltages, the trigger levels sometimes had to be increased significantly to avoid fake triggers, in rare cases up to 50\,mV, not because the Gaussian noise increased so much, but probably due to micro discharges. However, at high voltages also the signal was much higher due to the gain so that this could be afforded up to a certain limit. In general, the measurements were stopped at a voltage when the trigger level would have been needed to be set too high so that it would have cut into the signal distribution, or when the waveforms became instable, deformed or started not to return to the baseline, \eg due to micro discharges. For some devices, the last measured voltage point presented here is already affected by such instable waveforms, see \eg figure~\ref{fig:waveforms} right (typically one device started slightly earlier than the other), but in general the measurement was stopped then. An incident occurred during the measurement of the devices at the highest fluence of $10^{15}$\,n$_{eq}$/cm$^2$, when both sensors stopped working during a longer no-beam waiting period biased at 600\,V while being cooled with dry ice. Possible explanations include warming up and thermal run-away during that time. After that incident, both devices showed a reduced break down voltage of less than 1\,V and could not be used anymore for measurements.

\subsection{Measured waveform parameters and properties}
\label{sec:waveformParameters}

\begin{figure}[hbtp]
	\centering
	\includegraphics[width=7.5cm]{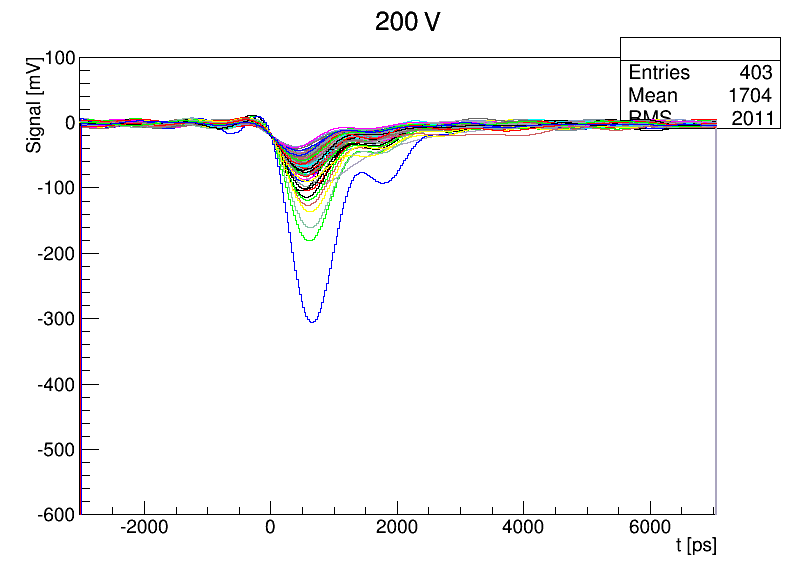}
	\includegraphics[width=7.5cm]{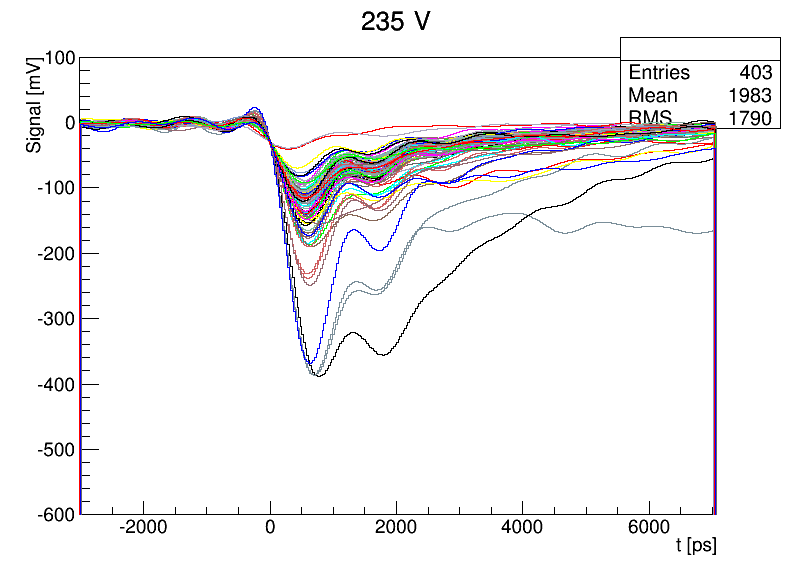}
	\caption{Examples of LGAD waveforms for med,unirr,L1 at 200\,V (left) and 235\,V (right).}
	\label{fig:waveforms}
\end{figure}

\begin{figure}[hbtp]
	\centering
	\includegraphics[width=7.5cm]{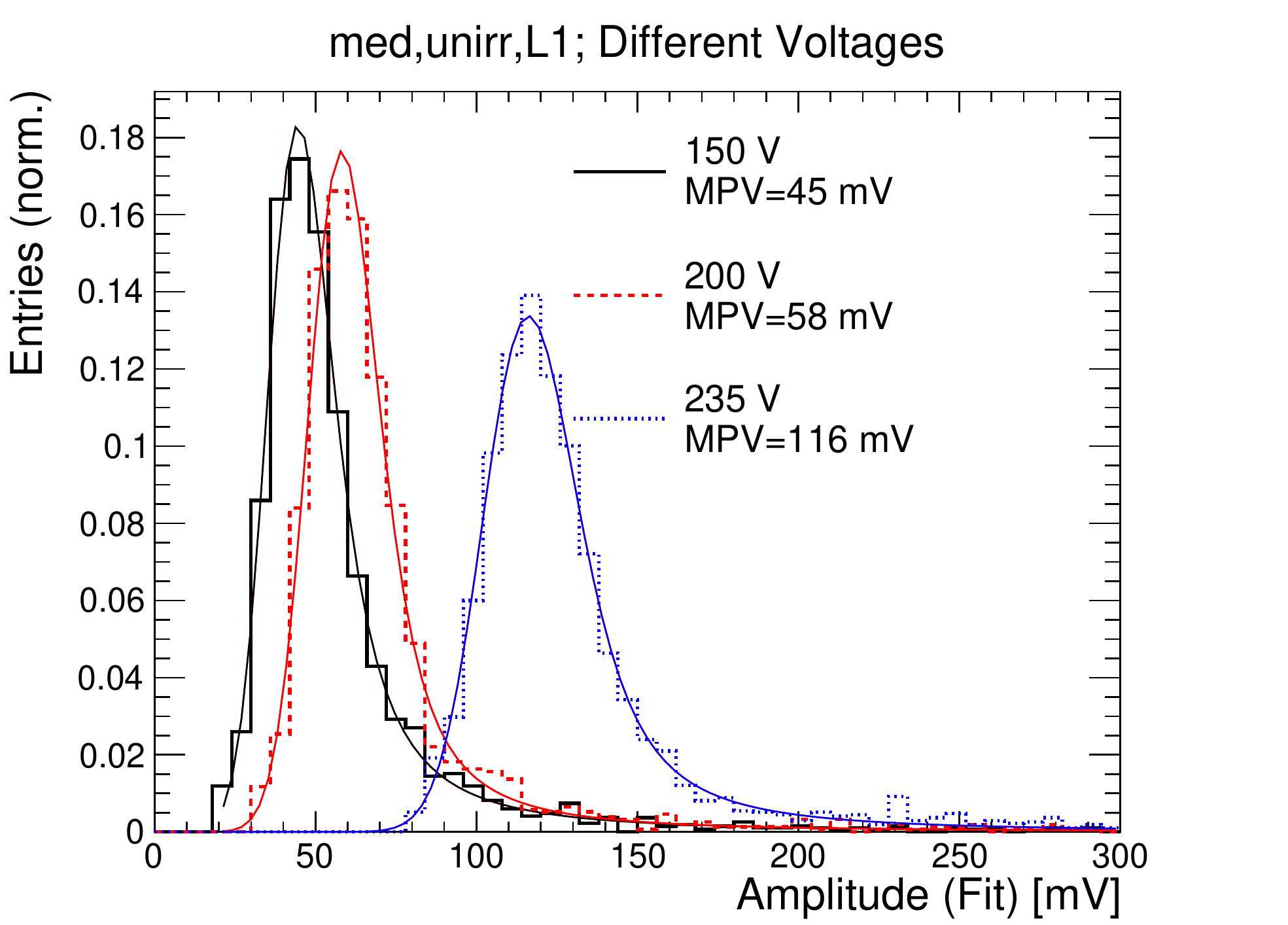}
	\includegraphics[width=7.5cm]{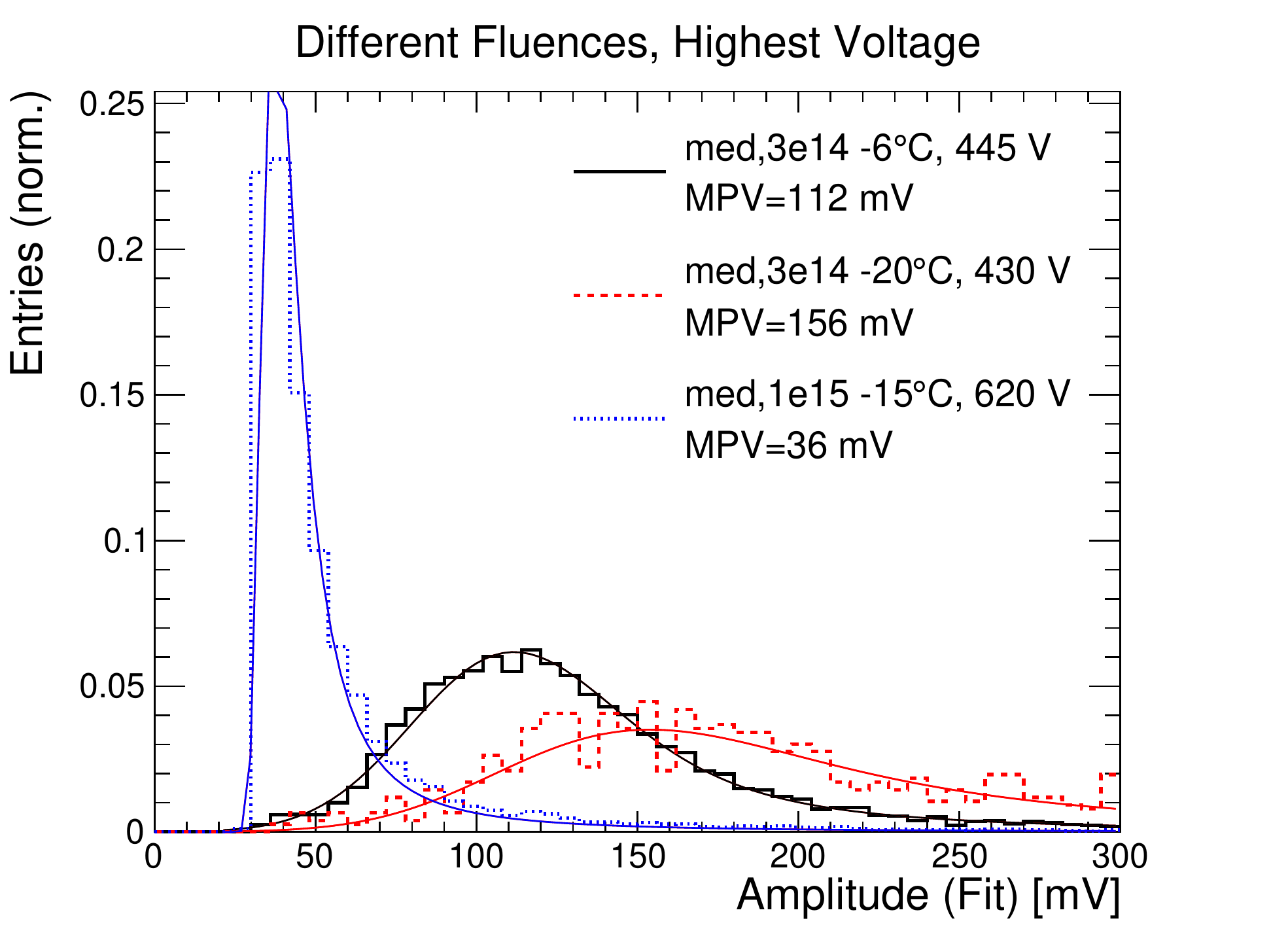}
	\caption{Amplitude distribution for med,unirr,L1 at different voltages (left) and for irradiated devices at the highest voltage measured (right). 
	}
	\label{fig:amplitude}
\end{figure}

\begin{figure}[hbtp]
	\centering
	\includegraphics[width=7.5cm]{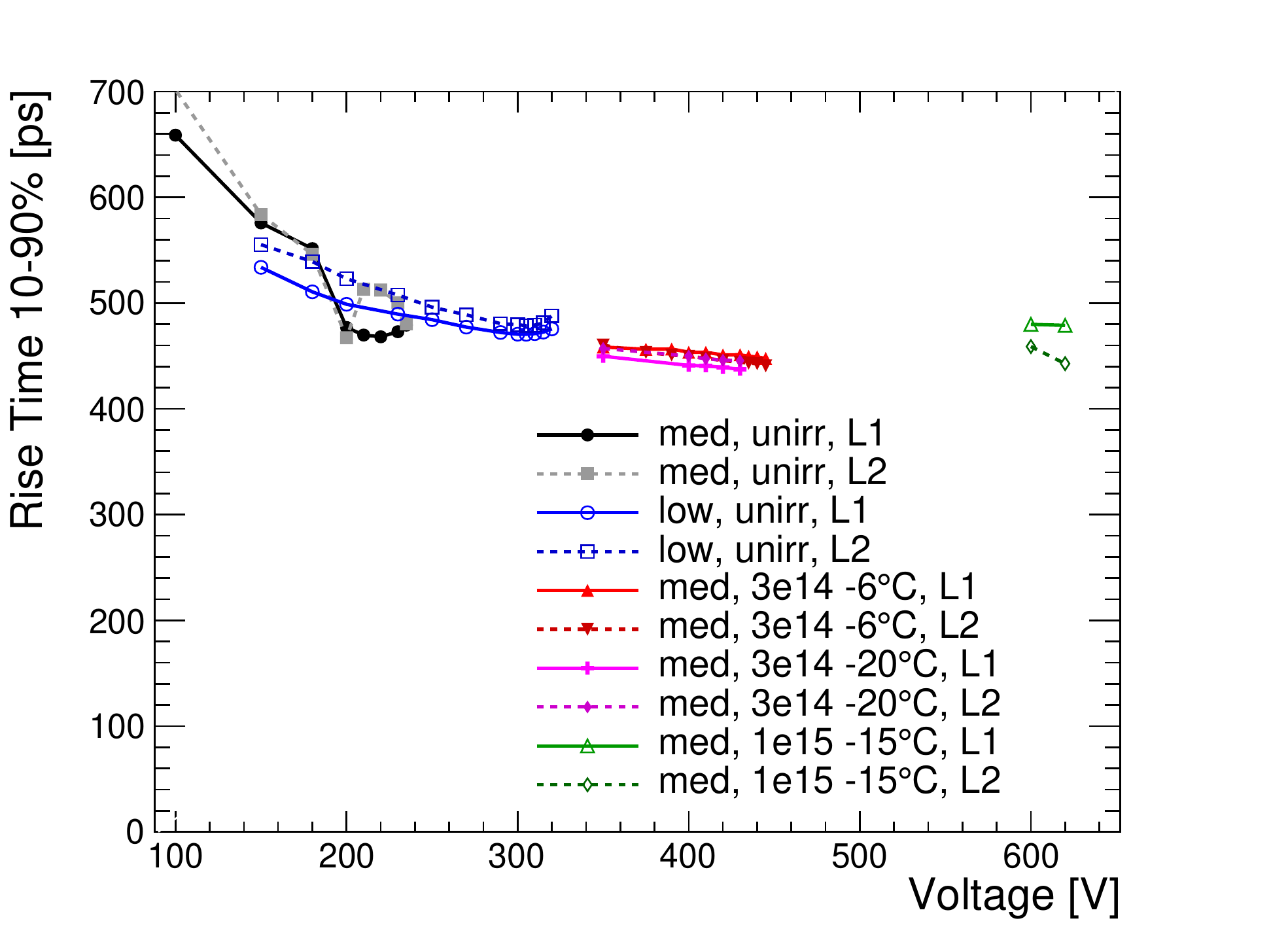}
	\includegraphics[width=7.5cm]{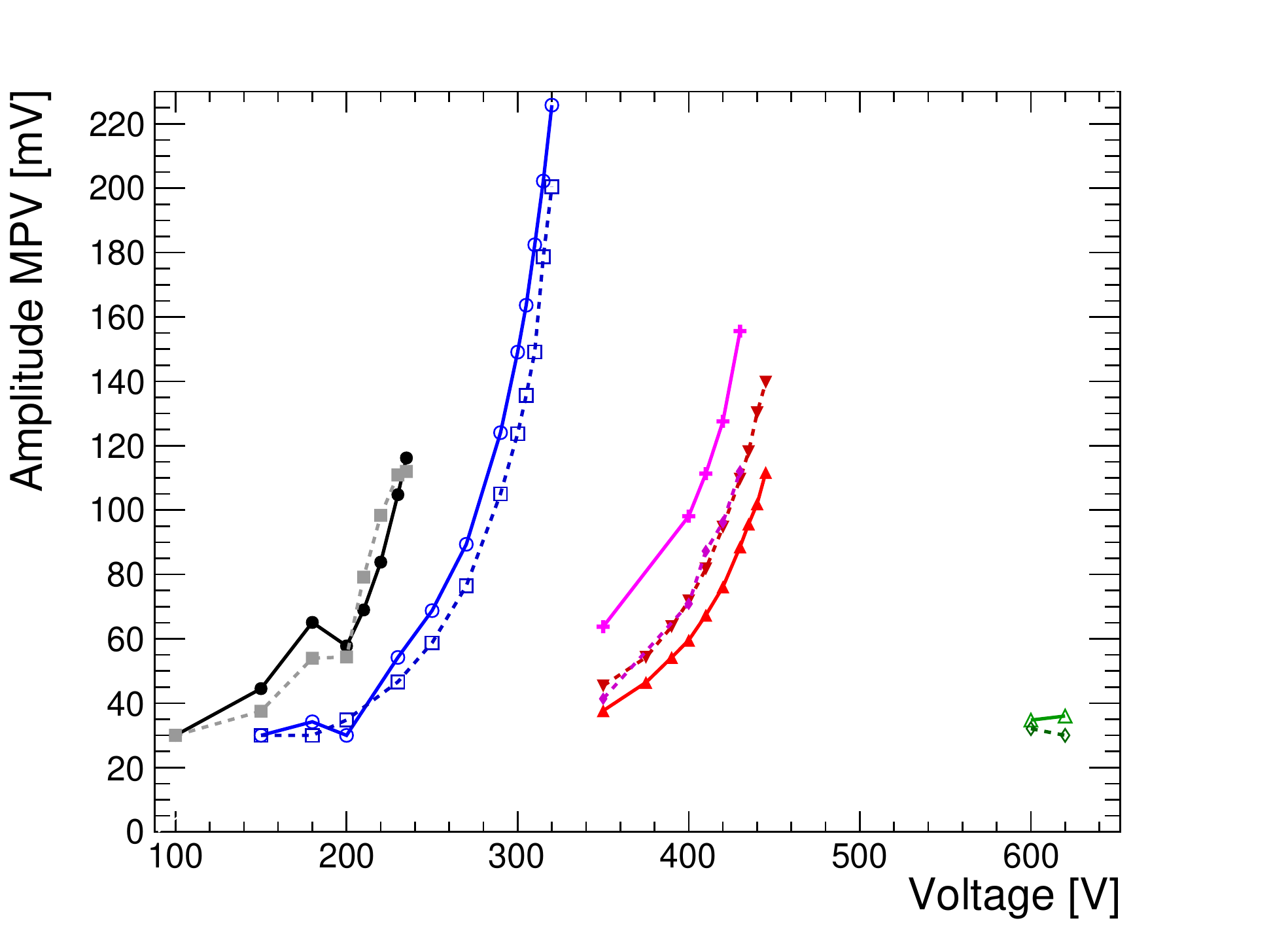}
	~
	\includegraphics[width=7.5cm]{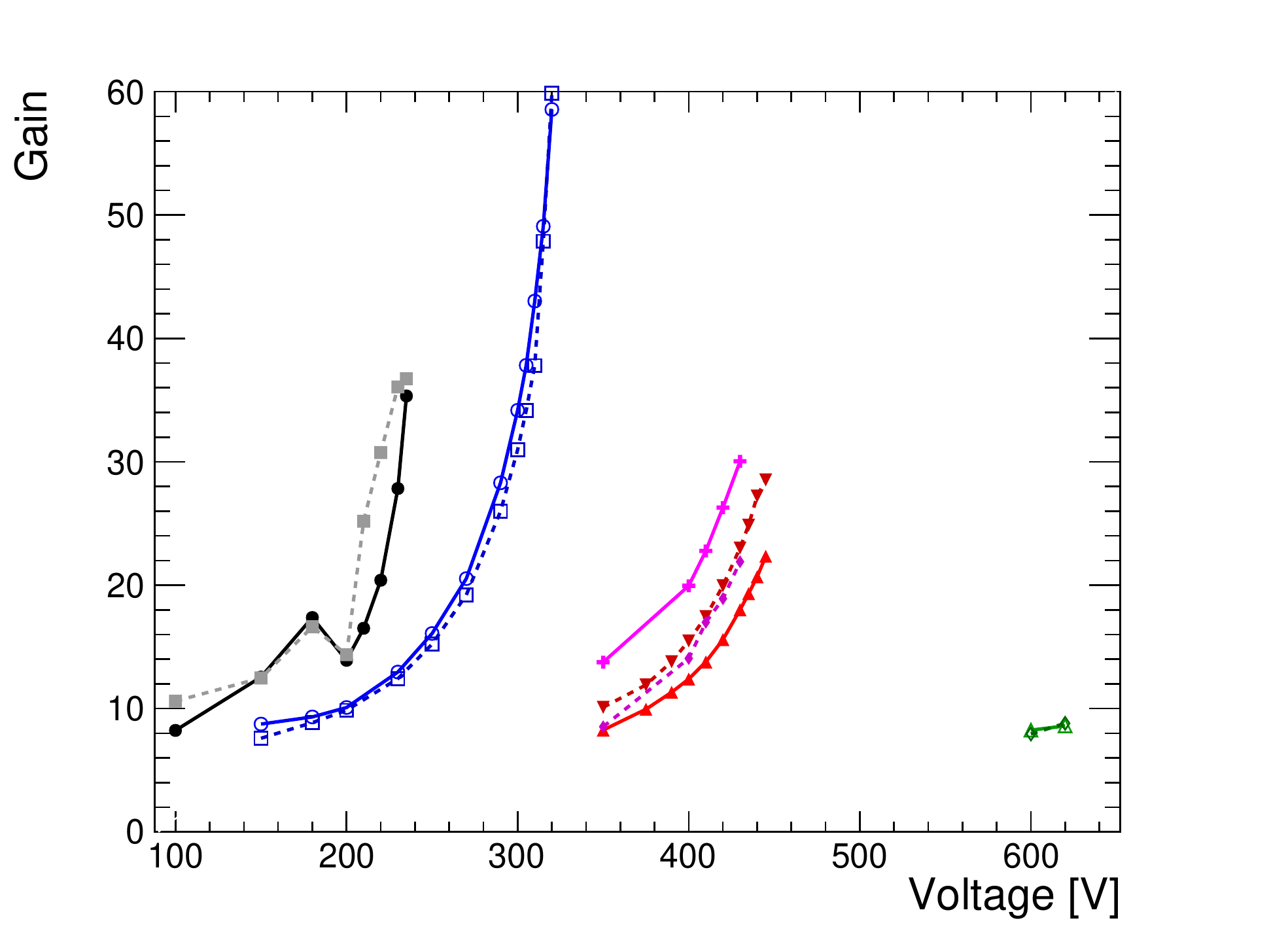}
	\includegraphics[width=7.5cm]{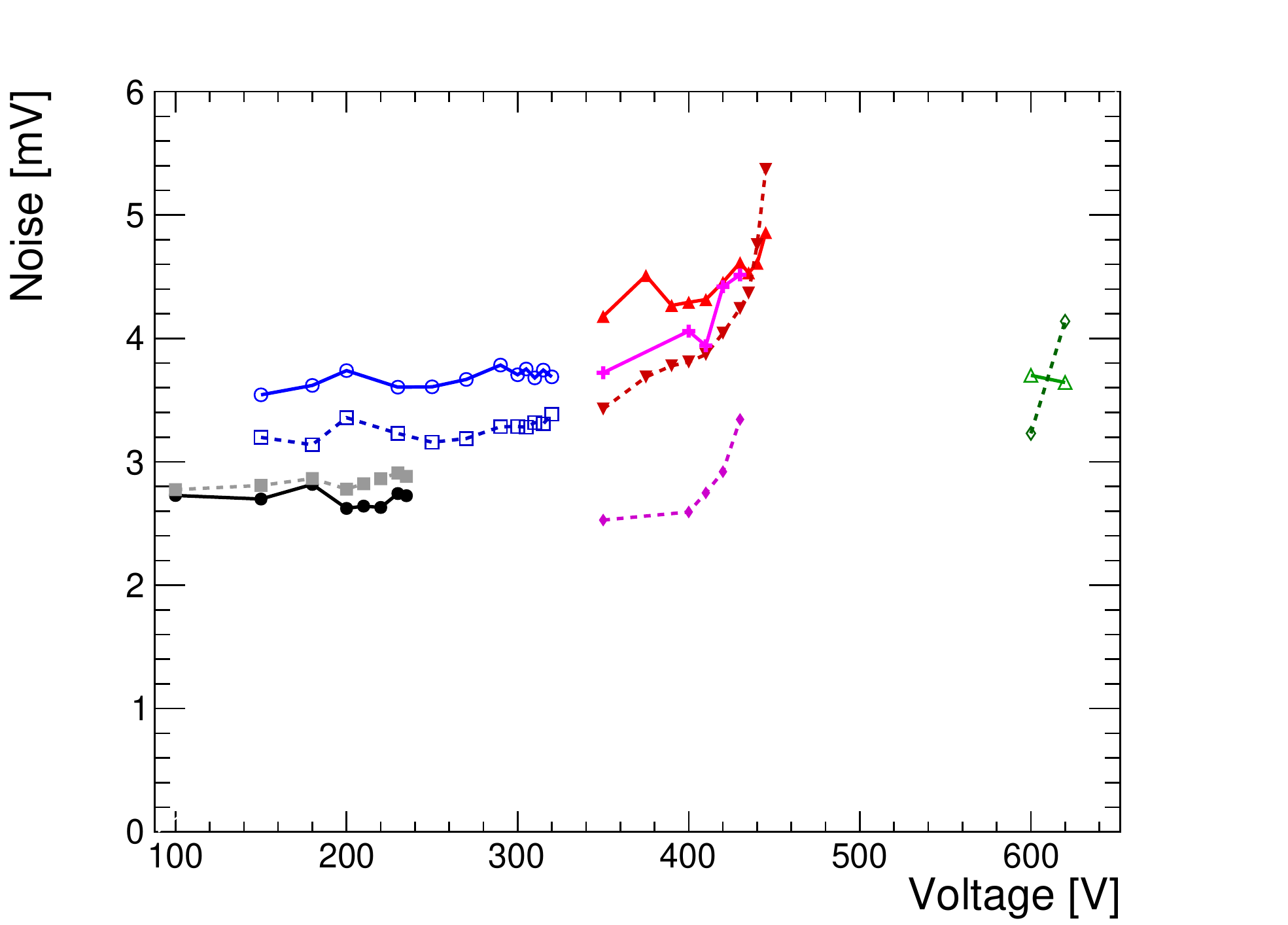}
	\includegraphics[width=7.5cm]{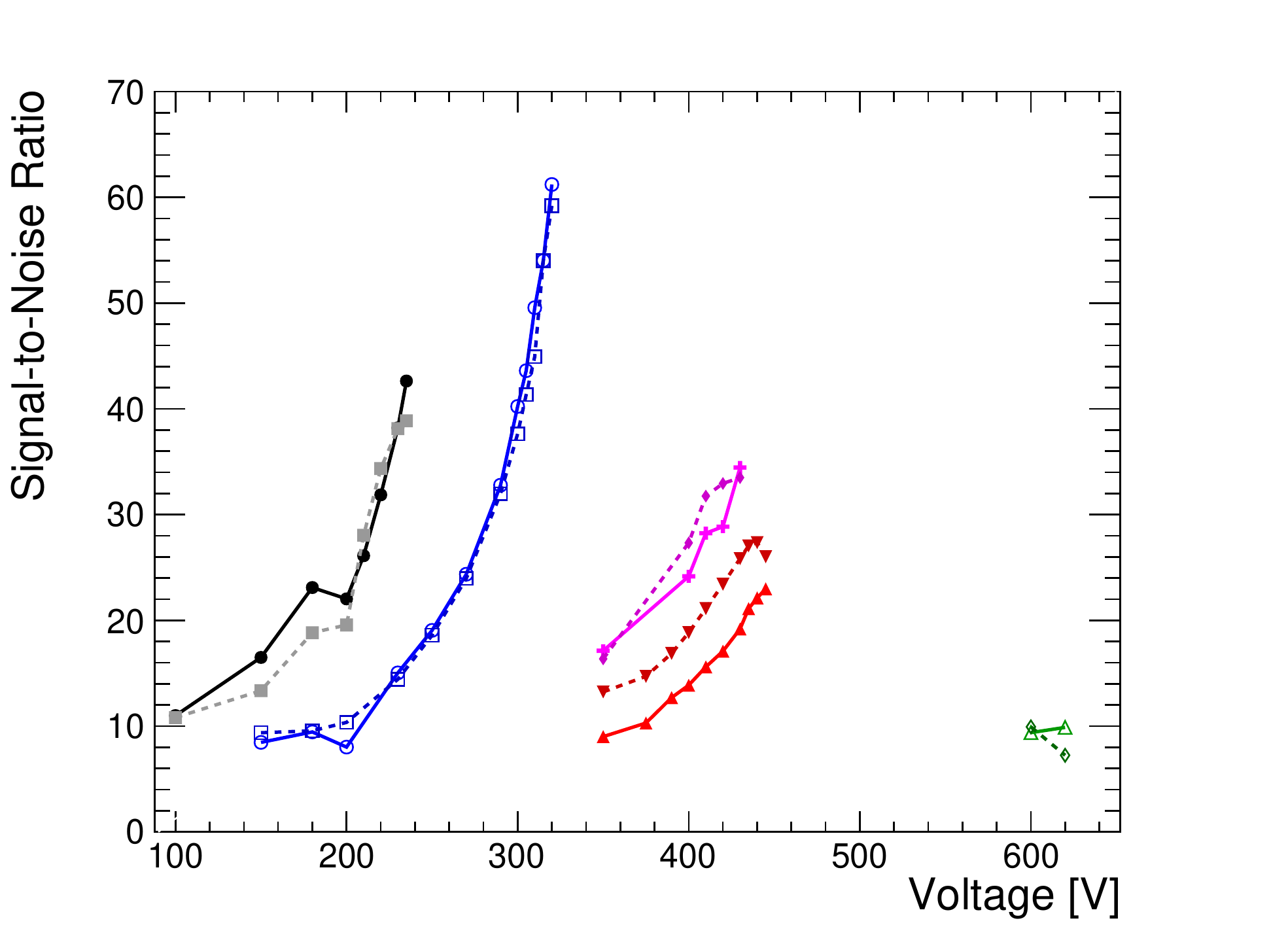}
	\includegraphics[width=7.5cm]{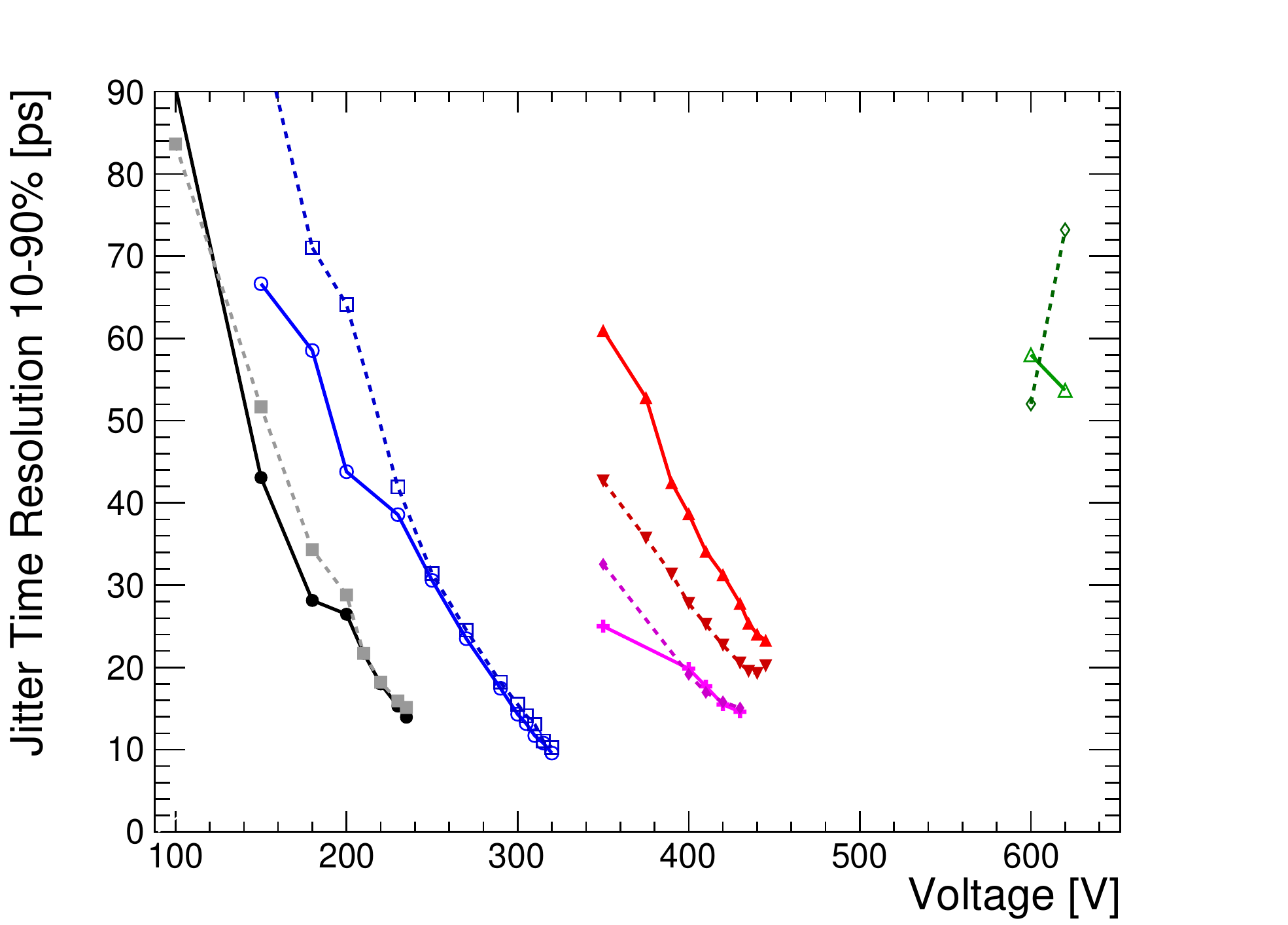}
	\caption{Overview on waveform parameters for all devices as a function of voltage.}
	\label{fig:parameterPlots}
\end{figure}

In this section, the measured waveforms are presented and their properties discussed. Figure~\ref{fig:waveforms} shows example waveforms at two voltages, and figure~\ref{fig:amplitude} the corresponding amplitude distributions for a few bias voltages. Figure~\ref{fig:parameterPlots} displays the voltage dependence of key waveform parameters, namely the rise time $\tau_{10-90\%}$\footnote{Measured as the time from 10 to 90\% of the amplitude.}, the most probable value (MPV) of the amplitude, the gain, the baseline noise N, the signal-to-noise ratio S/N and the jitter $\sigma_{jitter}$.

The waveforms are relatively fast and short with measured rise times of about 500\,ps and durations of about 1.5\,ns at high voltages (excluding the second peak about 1\,ns after the first one, which is believed to originate from artifacts of the setup such as impedance mismatch, since it is not observed in other setups~\cite{bib:UFSD50umTBNicolo,bib:HGTD}). The intrinsic waveform shape (before distortions from electronics) is a superposition of the induced currents from the drift of the primary electrons and holes and the secondary holes (see reference~\cite{bib:LGADDesignOpt} for a detailed simulation of the signal). The latter are created by impact ionisation in the charge multiplication layer from electrons that reach the front side (the field necessary for multiplication from holes is much higher than for electrons). This multiplication increases the number of charge carriers and hence the induced current as long as the primary electrons are drifting to the front side, which is expected to last for about 450\,ps at drift velocity saturation. This constitutes the intrinsic rise time of the signal, consistent with the measurement. As can be seen from figure~\ref{fig:parameterPlots}, the rise time slightly decreases with voltage due to the faster drift. Subsequently, the multiplied holes drift to the back side, which is expected to last slightly longer due to the slower drift of the holes, which adds to the total pulse width and gives a trapezoidal shape to the waveform. The intrinsic LGAD waveform is convoluted with the electronics response function that includes contributions from the sensor capacitance in combination with the 50\,$\Omega$ impedance of the amplifier with a rise time of $2.2\tau_{RC}=430$\,ps for a step function, and from the 1\,GHz band width of the oscilloscope-amplifier system with a rise time of 340\,ps for a step function. 

The amplitude distributions are shown in figure~\ref{fig:amplitude} for selected voltages and fitted with a Landau-Gauss function to extract the MPV. From figure~\ref{fig:parameterPlots} the strong voltage dependence of the MPV can be appreciated. A similar behaviour is obtained for the gain, which is extracted here by integrating the waveform from -1 to 4\,ns and dividing by the expected no-gain signal of 3.2\,ke$^{-}$, which is slightly different for 120\,GeV pions than for Sr90 beta particles~\cite{bib:PDG}. The signal-to-noise ratio was not high enough to directly measure the charge of a 45\,$\mu$m no-gain reference sensor, but the charge calibration has been verified with a 300\,$\mu$m thick pad diode. The in-situ gain from the beam test agrees with the Sr90 measurements from section~\ref{sec:lab} within the rather large run-to-run variations of 30--40\% mentioned in section~\ref{sec:setup}. The kink in the curves of med,unirr,L1/2 at 200\,V stems from such variations since measurements below and above have been obtained in different runs. Note that during the beam test measurements, it was possible to measure at slightly higher voltages than during the Sr90 tests, as mentioned before, and hence higher gains were achieved. In particular, the maximum gain of low,unirr was higher than for med,unirr, and the maximum gain of med,3e14 was approaching the one of med,unirr. Only for med,1e15, it was not possible to apply the maximum voltage from the Sr90 measurements, since the devices broke before as discussed above. The noise, as obtained from sampling the baseline before the signal at -1\,ns, was measured to be typically in the range between 3--4\,mV before irradiation and also after irradiation for low voltages, which is included by above mentioned run-to-run variations. However, whereas the unirradiated devices show no voltage dependence in the range measured, it is interesting to note that the noise of the irradiated devices increases with voltage, up to 5\,mV, probably due to increased shot noise from higher leakage currents. The signal-to-noise ratio, obtained as the ratio of the amplitude MPV and the noise, gave values as high as 60 before irradiation and up to 35 (10) at $3\times10^{14}$\,n$_{eq}$/cm$^2$ ($10^{15}$\,n$_{eq}$/cm$^2$). The time resolution contribution due to the electronics jitter can be estimated from the slope $dV/dt$ of the signal and the noise $N$ as $\sigma_{jitter} = N/(dV/dt) \approx \tau_{10-90\%}/(0.8\cdot S/N)$. It is steeply decreasing with voltage and reaches values as low as 10--15\,ps before irradiation and not much worse values of 15 (20)\,ps at $3\times10^{14}$\,n$_{eq}$/cm$^2$ measured at -20$^{\circ}C$ (-6$^{\circ}C$). However, for $10^{15}$\,n$_{eq}$/cm$^2$, the jitter is limited to 50\,ps.

\subsection{Time resolution}

\begin{figure}[hbtp]
	\centering
	\includegraphics[width=7.5cm]{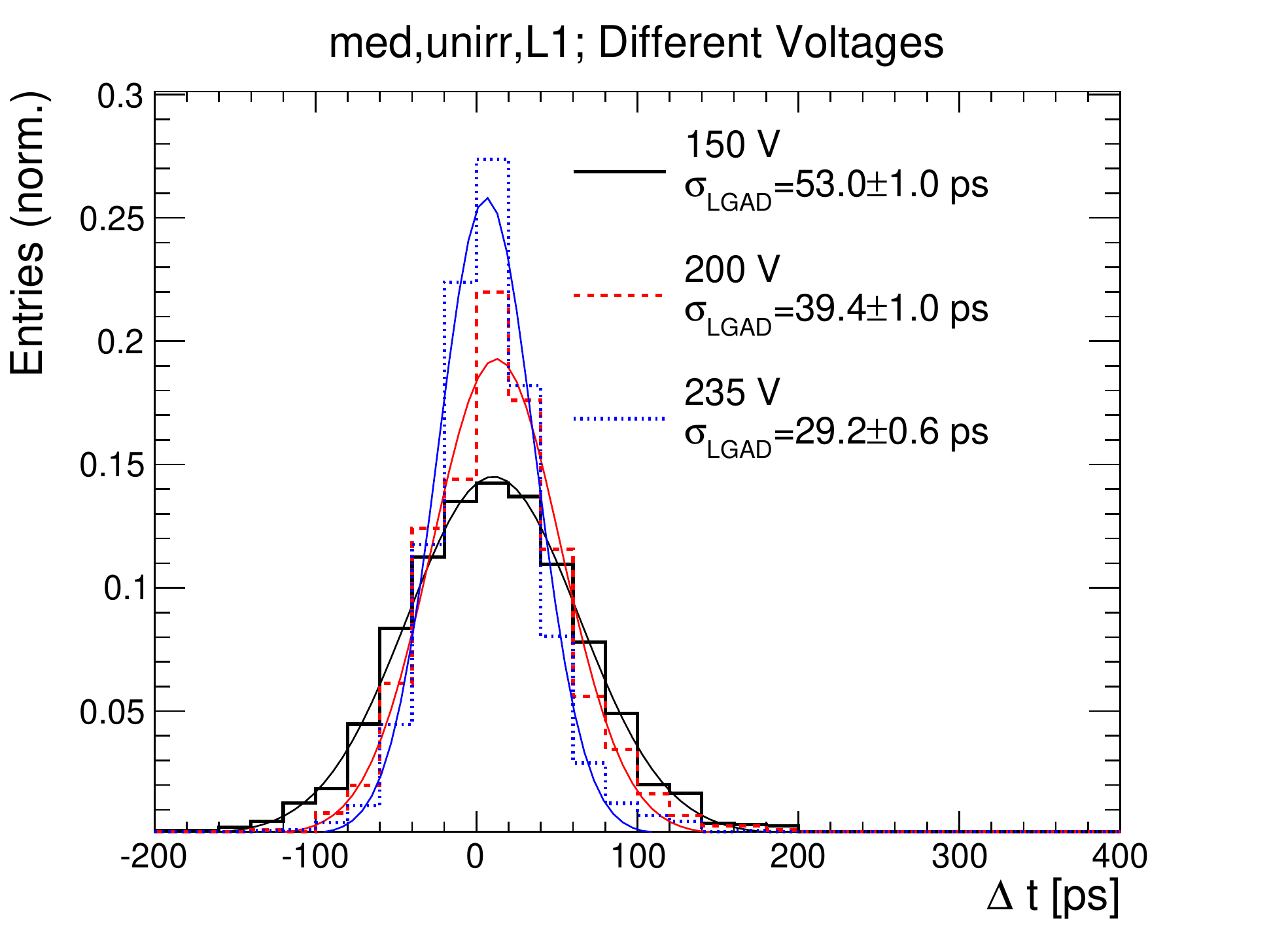}
	\includegraphics[width=7.5cm]{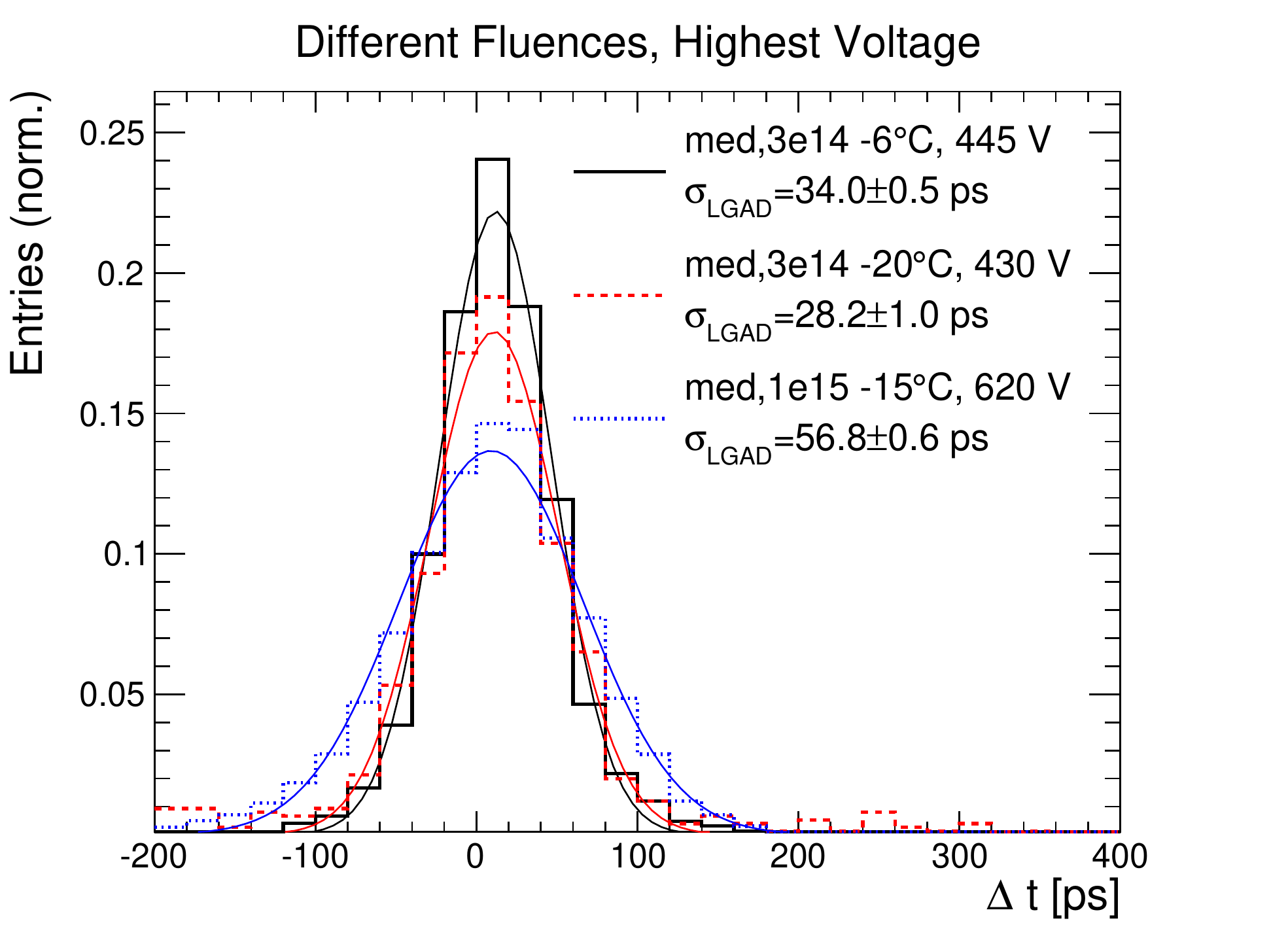}
	\caption{The time-of-arrival difference $\Delta t$ between typically an LGAD and SiPM2 for med,unirr,L1 at different voltages (left) and for irradiated devices at the highest voltage measured (right). The LGAD time resolution $\sigma_{LGAD}$ is obtained from a Gaussian fit after subtraction of the SiPM2 contribution. Only for med,3e14, at -20$^{\circ}$C, the $\Delta t$ shown is instead the difference between the two LGAD devices and $\sigma_{LGAD}$ is obtained from $\sigma_{total}/\sqrt{2}$.}
	\label{fig:timeResolutionDistributions}
\end{figure}

\begin{figure}[hbtp]
	\centering
	\includegraphics[width=7.5cm]{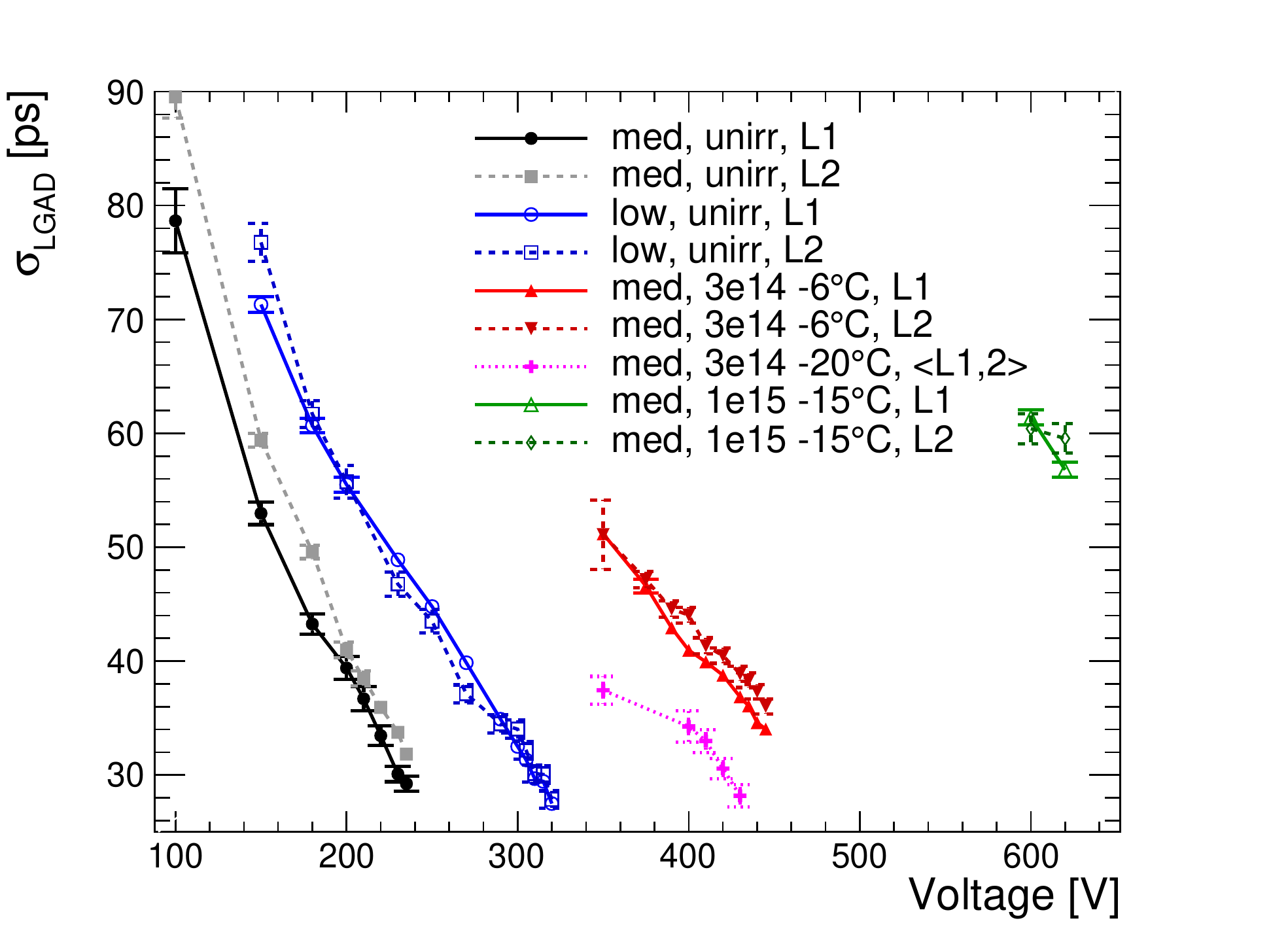}
	\includegraphics[width=7.5cm]{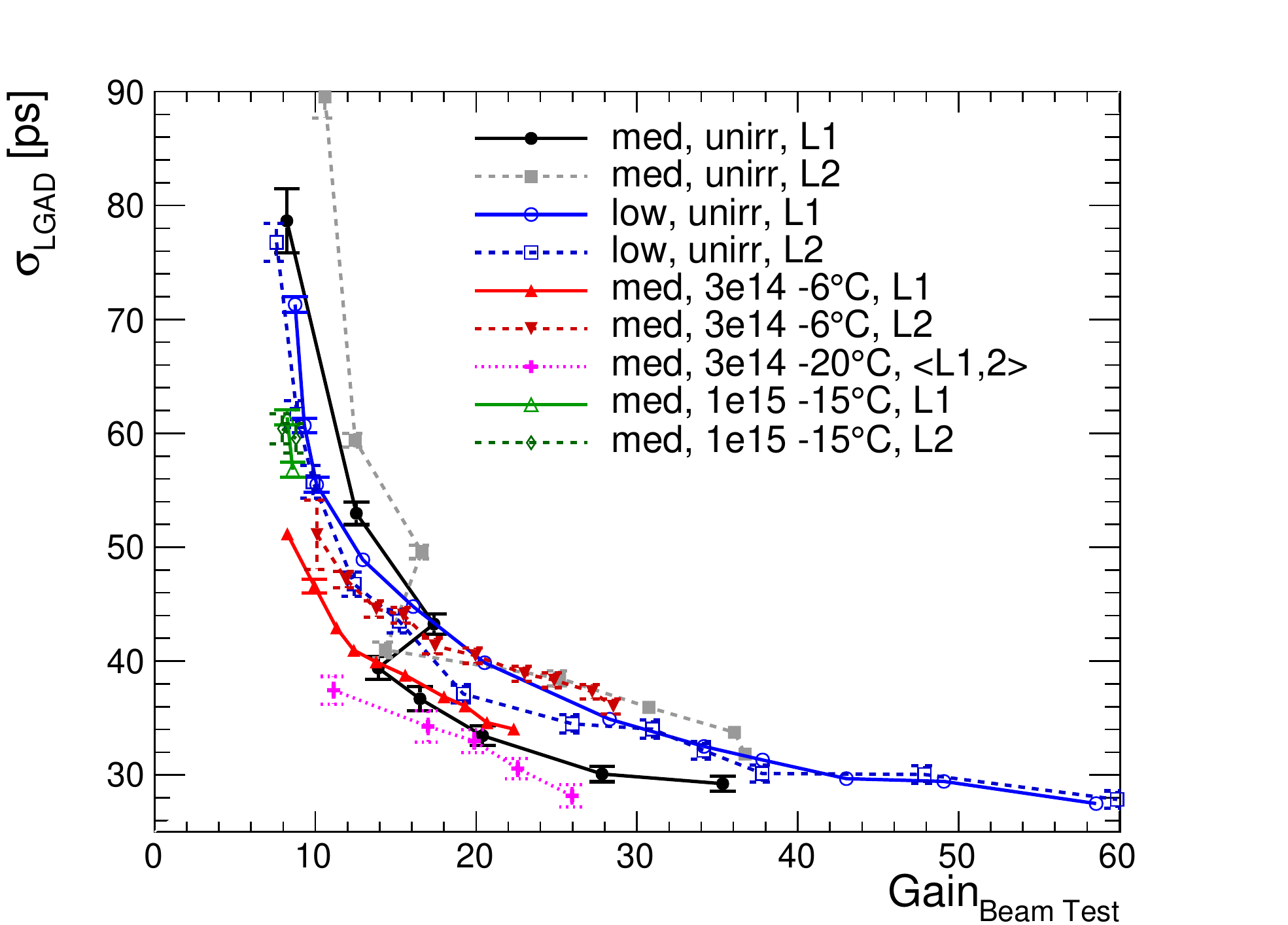}
	\caption{The LGAD time resolution $\sigma_{LGAD}$ as a function of voltage (left) and gain (right) for all devices. The gain has been obtained in-situ from the beam test measurement.}
	\label{fig:timeResolutionOverview}
\end{figure}

The time resolution was obtained from the spread of the time-of-arrival difference $\Delta t$ between two devices. This technique includes the contributions of both channels. For the digital CFD signals of the SiPMs, the time-of-arrival was simply taken at a fixed threshold of -250\,mV, which was about half of the constant amplitude. Variations of the threshold or more sophisticated algorithms were not found to change the results. For the analog LGAD signals without CFD, however, the time-of-arrival was obtained from an offline constant-fraction algorithm, which is needed to correct for the signal time walk due to amplitude variations. A threshold was set for each waveform individually at a certain fixed fraction of the amplitude, and the time of the threshold passing was interpolated linearly from the two measurement points just above and below that threshold. Many variations were tried such as including more points in a linear or polynomial fit or a spline interpolation, but no significant improvements in resolution were obtained, so that this simple and robust approach was taken. The fraction was scanned in 5\% steps from 10 to 90\% of the amplitude, and the optimal value was taken. Interestingly, this differed systematically for different voltages and devices, probably due to waveform shape variations that led to different points of the steepest slope $dV/dt$. For example, for the unirradiated LGAD devices, the optimal fraction was found to be around 80\% at low voltages, decreasing to 20\% at the highest measured voltages, whereas for the irradiated devices it stayed constant between 80 and 90\%.

Figure~\ref{fig:timeResolutionDistributions} shows example distributions of the time-of-arrival difference $\Delta t$ for different devices. 
The histogram spreads were extracted as the standard deviation of a Gaussian fit $\sigma_{total}$, which contains the contributions of both devices. The individual resolutions of the SiPM reference devices were determined from some runs at high LGAD voltages, in which two LGADs and two SiPMs were included in the data taking. In a combined analysis of all channel combinations, which allowed to disentangle the individual resolutions, they were determined as $\sigma_{SiPM1}=13\pm1$\,ps and $\sigma_{SiPM2}=7\pm1$\,ps, with the uncertainties containing statistical uncertainties of about 0.2\,ps and run-to-run variations of about 1\,ps. In the following, the LGAD resolutions $\sigma_{LGAD}$ of each individual LGAD device were determined by default from $\Delta t(LGAD-SiPM2)$ by subtracting quadratically $\sigma_{SiPM2}$, or, in case only two LGADs were measured as for med,3e14,L1/2 at -20$^{\circ}$C, the average resolution of both devices was obtained as $\sigma_{<LGAD>}=\sigma_{total}/\sqrt{2}$.

Figure~\ref{fig:timeResolutionOverview} (left) shows the voltage dependence of $\sigma_{LGAD}$ obtained in such a way for all devices measured. The uncertainties shown are statistical only (typically between 0.5--1\,ps). As mentioned above, run-to-run variations add a systematic uncertainty of about 10\%, \ie about 3\,ps at the highest voltages for each devices. The strong, almost linear decrease of $\sigma_{LGAD}$ with voltage can be seen. The two devices of the same type behave similarly, but for the different doses and fluences, there is an almost constant offset, mostly due to the different gain at a fixed voltage. In fact, if $\sigma_{LGAD}$ is plotted as a function of the gain as shown in figure~\ref{fig:timeResolutionOverview} (right), an approximately universal behaviour for all doses and fluences is observed. The spread of this curve is within the systematic uncertainties of the gain obtained in-situ in the beam test measurements of about 30--40\% (as discussed in section~\ref{sec:setup}). Slight deviations from a universal gain dependence are only expected if the noise shows significant variations, or if the drift velocities and hence rise times are very different for the same gain, \eg due to different voltages below saturation or different temperatures. 

The end-point time resolutions at the highest voltages measured are very similar for the medium and low doses before irradiation with 29\,ps at 235\,V and 28\,ps at 320\,V, respectively. The values before irradiation agree within the 10\% systematic uncertainties with the ones measured by other groups with a different setup~\cite{bib:UFSD50umTBNicolo,bib:HGTD}. 

Even after irradiation to $3\times10^{14}$\,n$_{eq}$/cm$^2$, the same time resolution as before irradiation could be obtained when measured at -20$^{\circ}$C, but at a higher voltage of 430\,V. When measured at -6$^{\circ}$C, the time resolution was about 8\,ps worse at the same voltage, which might be partly explained by a lower gain due to a lower impact ionisation coefficient at higher temperatures, and partly due to lower drift velocities. A more detailed study of the temperature dependence, also before irradiation, will be presented in a later paper. 

Whereas at $3\times10^{14}$\,n$_{eq}$/cm$^2$, similar time resolutions could be achieved as before irradiation, the voltage stability of the devices irradiated to $10^{15}$\,n$_{eq}$/cm$^2$ was not good enough to compensate for the effective acceptor removal in the multiplication layer. The resolution at the highest measured voltage of 620\,V was found to be minimally 57\,ps. 

Better time resolutions might be achievable with more sophisticated algorithms that exploit the full waveform information. However, for HEP applications with a very large number of channels, a simple solution such as a CFD followed by a TDC is the most realistic option. This is why the above analysis with the offline constant-fraction algorithm tried to stay as close as possible to that scenario. Furthermore, for one run with a device of the medium dose before irradiation, an AFP CFD was inserted between the amplifier and the oscilloscope with a constant-fraction parameter of 50\% of the amplitude and a threshold below which events were rejected of 50\,mV (hence only high voltages could be measured). At a voltage of 230\,V, a time resolution of 35\,ps was achieved, which is only slightly worse than the 30\,ps achieved without CFD at the same voltage, especially considering the 10\% systematic uncertainty and the non-optimised CFD threshold and fraction for this first test. 

\section{Conclusions and outlook}
\label{sec:conclusions}

Low Gain Avalanche Detectors (LGADs) were produced with an active thickness of about 45\,$\mu$m, and their gain and time resolution were studied for the first time for different initial multiplication layer implantation doses and before and after irradiation with neutrons up to $10^{15}$\,n$_{eq}$/cm$^2$.

The gain showed the expected decrease at a fixed voltage for a lower implantation dose and a higher fluence due to lower acceptor concentrations in the multiplication layer. Time resolutions below 30\,ps were obtained at the highest applied voltages for both doses before irradiation, as well as after a fluence of $3\times10^{14}$\,n$_{eq}$/cm$^2$. Also, the time resolution was found to follow approximately a universal function of gain for all devices. This shows that given a good voltage stability that allows to reach a certain gain, different devices can reach similar time resolutions. However, at $10^{15}$\,n$_{eq}$/cm$^2$, the time resolution at the maximum applicable voltage of 620\,V during the beam test was degraded to 57\,ps since the voltage stability was not good enough to compensate for the doping loss in the multiplication layer. Further investigations on the voltage stability and dependence on environmental conditions such as temperature and humidity are envisaged, as well as the addition of more fluence steps and other irradiation types.

These results demonstrate that thin LGADs are good candidates for HEP experiments with 10--30\,ps timing requirements, since the one-layer time resolution obtained here can be improved using multiple layers as demonstrated \eg in reference~\cite{bib:UFSD50umTBNicolo}. LGADs are an option for experiments with moderate radiation levels of at least up to $3\times10^{14}$\,n$_{eq}$/cm$^2$. Limitations can arise from the observed resolution degradation at about $10^{15}$\,n$_{eq}$/cm$^2$. Also here, the same methods of using multiple layers could help to bring back a part of the performance, \eg still about 30 (20)\,ps could be reached at that fluence using 4 (9) layers. Forward experiments like AFP or CT-PPS have the additional complication of non-uniform irradiation, which would make it necessary to apply different bias voltages to different parts of the detector to achieve a homogeneous timing performance throughout the whole device after irradiation.

Hence, although the first results on irradiated LGADs obtained here are promising, eventually it would be desirable to make the LGAD technology intrinsically more radiation hard. Possible solutions being investigated include Gallium instead of Boron doping for the p-type multiplication layer or Carbon enhancement~\cite{bib:GregorTrento}. First devices with these modifications have been produced at CNM and are under study~\cite{bib:MarTrento}. Furthermore, it is considered to investigate devices with even thinner active area since on the one hand, in general a better time resolution is expected, and on the other hand the electric field and hence charge multiplication in the bulk region is enhanced, especially after irradiation, making them potentially more radiation hard.

\acknowledgments 
This work was partly performed in the frameworks of the ATLAS Forward Proton and the CERN RD50 collaborations. The authors wish to thank the CERN-SPS NA team for excellent beam and infrastructure support; DESY and Hamburg University (especially H.\,Jansen and M.\,Matysek) for providing the PCBs; and M.\,Moll, the CERN SSD group, E.\,Griesmayer  and L.\,Paolozzi for providing and help with the amplifiers.

This work was partially funded by: the MINECO, Spanish Government, under grants FPA2013-48308-C2-1-P, FPA2014-55295-C3-2-R, FPA2015-69260-C3-2-R, FPA2015-69260-C3-3-R (co-financed with the European Union's FEDER funds) and SEV-2012-0234 (Severo Ochoa excellence programme), as well as under the Juan de la Cierva programme; the Spanish ICTS Network MICRONANOFABS partially supported by MINECO; the Catalan Government (AGAUR): Grups de Recerca Consolidats (SGR 2014 1177); the Czech programmes IGA\_Prf\_2017\_005 of Palacky University and MSMT INGO II \v{c}. LG15052; and the European Union's Horizon 2020 Research and Innovation programme under Grant Agreement no. 654168 (AIDA-2020).



\providecommand{\href}[2]{#2}\begingroup\raggedright\endgroup

\end{document}